\documentclass[psfig]{aa}
\begin{document}

\thesaurus{16 (03.13.4 Methods: numerical; 02.18.7 Radiative transfer;
08.01.3 Stars: atmospheres; 02.12.1 Line: formation)}

\title{Monte Carlo transition probabilities}


\author{L.B. Lucy}

\offprints{L.B. Lucy}

\institute{Astrophysics Group, Blackett Laboratory, Imperial College of
Science, Technology and Medicine, Prince Consort Road, London SW7~2BW}

\date{Received ; accepted }

\maketitle

\begin{abstract}
Transition probabilities governing the interaction of energy packets and 
matter
are derived that allow Monte Carlo NLTE transfer
codes to be constructed without simplifying the treatment of line formation.
These probabilities are such that the Monte Carlo calculation
asymptotically recovers the local emissivity of a gas in statistical 
equilibrium. Numerical experiments with one-point statistical equilibrium
problems for Fe II and Hydrogen confirm this asymptotic behaviour.
In addition, the resulting Monte Carlo emissivities are shown
to be far less sensitive to errors in the populations of the emitting 
levels than are the values obtained with the basic emissivity formula.


\end{abstract}

\section{Introduction}

When Monte Carlo methods are used to compute the spectra of astronomical
sources,
it is advantageous to work with {\em indivisible} monochromatic
packets of radiant energy
and to impose the constraint that, when interacting with matter, their
energy is conserved in the co-moving frame. The first of these
constraints leads to simple code and the second facilitates convergence to an
accurate temperature stratification.

	For a static atmosphere, the energy-conservation constraint
automatically gives a divergence-free radiative flux
even when the temperature
stratification differs from the radiative equilibrium solution. A remarkable
consequence is that the simple $\Lambda$-iteration
device of adjusting the temperature to bring the matter into thermal
equilibrium with the Monte Carlo radiation field results in rapid convergence
to the close neighbourhood of the radiative equilibrium solution (Lucy 1999a).
An especially notable
aspect of this success is that this temperature-correction procedure is
geometry-independent, and so these methods readily generalize to 2- and
3-D problems.    
	
	For an atmosphere in differential motion, the energy-conservation
constraint yields a 
radiative flux that is rigorously divergence-free in every local matter
frame. Determining the temperature stratification by bringing matter into
thermal equilibrium with such a radiation field - i.e., by imposing radiative
equilibrium in the co-moving frame - is an excellent approximation if the
local cooling time scale is short compared to the local expansion time scale.
This condition is well satisfied for the spectrum-forming layers of
supernovae (SNe) and of hot star winds (Klein \& Castor 1978).

        The constraint that the energy packets be indivisible is advantageous
from the point of view of coding simplicity. The
interaction histories of the packets are then followed one-by-one as they
propagate through the computational domain, with there being no necessity to
return to any of a packet's interactions in order to continue or complete
that interaction. This is to be contrasted with a Monte Carlo code that
directly simulates physical processes by taking its quanta to be a
sampling of the
individual photons. Absorption of a Monte Carlo quantum is then
often followed by the emission of several quanta as an atom cascades back to
its ground state. Multiple returns to this interaction are then
necessary in order to follow the subsequent paths of each of these cascade
quanta.
The resulting coding complexity is of course compounded by some of these 
quanta creating further cascades.  

	Although coding simplicity argues strongly for indivisible
packets, a counter argument is the apparent implied need to 
approximate the treatment of line formation. Thus, in Monte Carlo codes for
studying the dynamics of stellar winds (Abbott \& Lucy 1985; Lucy \& Abbott
1993) or for synthesizing the spectra of SNe (Lucy 1987; Mazzali \&
Lucy 1993), the integrity of the packets could readily be maintained since 
lines were assumed to form by coherent scattering in the 
matter frame. But significantly, an improved SN code has recently been
described (Lucy 1999b) in which branching into the alternative downward
transitions is properly taken into account without sacrificing
indivisibility. Accordingly, an obvious question now is whether Monte Carlo
techniques can be developed that enforce
energy-packet indivisibility and yet do not have to adopt {\em any}
simplifications with regard to
line formation. If this can be achieved, then
Monte Carlo
codes for general NLTE transfer problems become feasible.  

\section{Macro-atoms}

As discussed in Sect. 1, it is common in Monte Carlo transfer codes to
quantize radiation into monochromatic energy packets. But 
matter is not quantized, neither naturally into individual atoms 
nor artificially into parcels of matter. 
Instead, the
continuum description of matter is retained, with macroscopic absorption
and scattering coefficients governing the interaction histories of the energy
packets.

	Nevertheless, it now proves useful to imagine that matter is
quantized into
{\em macro-atoms} whose properties are such that their interactions with
energy
packets asymptotically reproduce the emissivity 
of a gas in statistical equilibrium.
But these macro-atoms, unlike energy packets, do not explicitly
appear in the Monte Carlo code. As conceptual constructs, they 
facilitate the derivation
and implementation of the Monte Carlo transition probabilities that allow
in an accurate treatment of line formation.	

	The general properties of macro-atoms are as follows:

	1) Each macro-atom has discrete internal states in one-to-one
correspondence with the energy levels of the atomic species being represented.

	2) An inactive macro-atom can be activated to one of its internal
states $i$ by absorbing a packet of kinetic energy or a packet  
of radiant energy of an appropriate co-moving frequency.

	3) An active macro-atom can undergo an internal transition from state
$i$ to any other state
$j$ without absorbing or emitting an energy packet.

	4) An active macro-atom becomes inactive by emitting
a packet of kinetic energy or a packet of radiant energy of an appropriate
co-moving frequency.

	5) The de-activating packet has the same energy in the
macro-atom's frame as the original activating packet.

\begin{figure}
\vspace{5.5cm}
\includegraphics{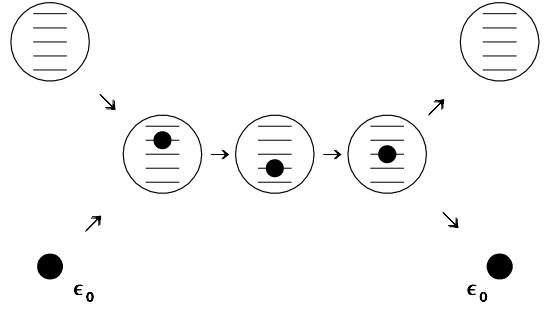}
\caption{Schematic representation of the interaction of a macro-atom with
a packet of energy $\epsilon_{0}$. The macro atom is activated by absorbing
the energy packet, makes two internal transitions, and then de-activates
by emitting a packet of energy $\epsilon_{0}$.  }

\end{figure}

	Figure 1 illustrates these general rules. An inactive
macro-atom, with internal states shown schematically, encounters
a packet of energy $\epsilon_{0}$ and is activated to one of these states.
The active macro-atom then undergoes two internal
transitions before de-activating itself by emitting a packet of
energy $\epsilon_{0}$.

	Subsequently, energy packets will in general be referred to as
$e$-packets but also 
as $r$- or $k$-packets when specifying their contents to be radiant or
kinetic energy, respectively.

\section{Transition probabilities}

	In Sect. 2, the concept of a macro-atom was introduced by stating
some general properties concerning its interaction with $e$-packets. 
The challenge now is to derive explicit rules governing
a macro-atom's activation, its subsequent internal transitions,
and its eventual de-activation. Asymptotically, the result of obeying these
rules must be the emissivity corresponding to statistical equilibrium.   

\subsection{Energy flow rates}

	For the moment, we drop the notion of a macro-atom and consider a
real atomic species interacting with its environment.
Let $\epsilon_{i}$ denote the excitation {\em plus} ionization energy of
level $i$ and let $R_{ij}$ denote the radiative rate for the
transition $i \rightarrow j$.
The rates per unit volume at which transitions into and out of $i$
absorb and emit radiant energy are then
\begin{equation}
 \dot{A}_{i}^{R} = R_{\ell i} \epsilon_{i \ell} \;\;\;\;\; and \;\;\;
\;\;  \dot{E}_{i}^{R} = R_{i \ell} \epsilon_{i \ell} \;\;\; ,
\end{equation}
respectively, where
$ \epsilon_{i \ell} = h \nu_{i \ell} = \epsilon_{i}-\epsilon_{\ell}$.
Note the summation convention adopted for the suffix $\ell$,
which ranges over all levels $<i$, including those of lower ions. Similarly,
below, the suffix $u$ implies summation over all levels $>i$, including those
of higher ions.

	The corresponding rates at which kinetic energy is absorbed from, or
contributed to, the thermal pool by transitions to and from level $i$ are  
\begin{equation}
 \dot{A}_{i}^{C} = C_{\ell i} \epsilon_{i \ell} \;\;\;\;\; and
\;\;\;
\;\;  \dot{E}_{i}^{C} = C_{i \ell} \epsilon_{i \ell} \;\;\; ,
\end{equation}
where $C_{ij}$ is the collisional rate per unit volume for the transition
$i \rightarrow j$.

	If we now define the total rate for the transition
$i \rightarrow j$ to be ${\cal R}_{ij} = R_{ij}+C_{ij}$, then the net rate at
which level $i$ absorbs energy is
\begin{equation}
 \dot{A}_{i}^{R}+\dot{A}_{i}^{C}-\dot{E}_{i}^{R}-\dot{E}_{i}^{C}=
 ({\cal R}_{\ell i}-{\cal R}_{i \ell})
 (\epsilon_{i}-\epsilon_{\ell}) \;\;\; .
\end{equation}
This is an identity that follows directly from the defining
Eqs.(1) and (2); it is therefore quite general and does not assume statistical
equilibrium.

\subsection{Statistical equilibrium}

We now assume that the level populations $n_{i}$ are in statistical
equilibrium. For level $i$, this implies that
\begin{equation}
 ({\cal R}_{\ell i}-{\cal R}_{i \ell})+({\cal R}_{ui}-{\cal R}_{iu})=0 \;\;\;.
\end{equation}

	A useful alternative representation of statistical equilibrium is 
obtained by multiplying Eq.(4) by $\epsilon_{i}$ and then eliminating the term
$({\cal R}_{\ell i}-{\cal R}_{i \ell})\epsilon_{i}$ using Eq.(3). The result
can be written in the form
\begin{eqnarray}
 \dot{E}_{i}^{R}+\dot{E}_{i}^{C}+{\cal R}_{iu}\epsilon_{i}+
 {\cal R}_{i \ell}\epsilon_{\ell}  \nonumber \\  
 = \dot{A}_{i}^{R}+\dot{A}_{i}^{C}+{\cal R}_{ui} \epsilon_{i}
 +{\cal R}_{\ell i} \epsilon_{\ell}            \;\;\; .       
\end{eqnarray}

	Eq.(4), the conventional equation of statistical equilibrium,
balances
the rates at which basic atomic processes excite and de-excite level $i$. As
such, it directly relates to Nature's quantization of radiation into photons
and of matter into atoms. In contrast, Eq.(5), though mathematically
equivalent,
deals with macroscopic energy flow rates in a finite volume element. These
flows can now be quantized into indivisible $e$-packets. Moreover,
we can think of the volume element as a macro-atom with discrete energy
states.   
\subsection{Interpretation}
Eq.(5) expresses the fact that in statistical equilibrium the
contribution from level $i$ to the energy 
content of unit volume is stationary. In consequence, the net rate at which
level $i$ gains energy - the right-hand side of Eq.(5) - equals the net
rate of loss - the left-hand side.

	But the importance here of Eq.(5) lies in the various terms
contributing
to gains and losses by level $i$ and their relevance for constucting
transition rules for macro-atoms.  The net rate of gain comprises the
expected absorption terms $\dot{A}_{i}^{R}$ and $\dot{A}_{i}^{C}$ plus
the terms ${\cal R}_{ui} \epsilon_{i}$ and ${\cal R}_{\ell i} \epsilon_{\ell}$
that clearly represent energy flows into $i$ from upper and lower levels.
Similarly, the net rate of loss comprises the expected emission
terms $\dot{E}_{i}^{R}$ and $\dot{E}_{i}^{C}$ plus the terms
${\cal R}_{iu}\epsilon_{i}$ and ${\cal R}_{i \ell}\epsilon_{\ell}$
representing energy flows out of $i$ to upper and lower levels.

	The above remarks imply definitive values for the energy
flows between level $i$ and other levels. But this is not true. If 
Eq.(4) is rewritten as
\begin{equation}
 {\cal R}_{iu}+{\cal R}_{i\ell}={\cal R}_{ui}+{\cal R}_{\ell i}  \;\;\; ,
\end{equation}
then comparison with Eq.(5) shows immediately that an arbitrary
quantity of energy $\epsilon$ may be added to $\epsilon_{i}$ and 
$\epsilon_{\ell}$ without invalidating this equation. But this
merely shifts the zero point of the energy scale for excitation and
ionization, which we are always free to do. Nevertheless, this freedom
implies a corresponding indefiniteness in the energy flow
rates between levels.  

\subsection{Stochastic interpretation}
Notwithstanding this indefiniteness, we now interpret Eq.(5) in terms
of macro-atoms absorbing and emitting $e$-packets or undergoing
transitions between internal states.
In this interpretation, the terms $\dot{A}_{i}^{R}$ and
$\dot{A}_{i}^{C}$ obviously represent the activation of macro-atoms to
state $i$ due
to the absorption of $r$-packets and of $k$-packets,
respectively.

	Now consider an ensemble of active macro-atoms in state $i$. For this
ensemble to reproduce the behaviour of the real system, the
relative numbers of the macro-atoms that subsequently
de-activate with the
emission an $r$- or $k$-packet or which make a
transition to another internal state must be proportional to the relative
values of
the corresponding 
terms on the left-hand side of Eq.(5). Accordingly, for an individual
macro-atom in state $i$, the probabilities that it de-activates
with the emission of an $r$-packet or a $k$-packet are
\begin{equation}
 p_{i}^{R}= \dot{E}_{i}^{R}/D_{i} \;\;\;\;\; and \;\;\;\;\;
             p_{i}^{C}=\dot{E}_{i}^{C}/D_{i} \;\;\; ,
\end{equation}
where  
\begin{equation}
 D_{i}= \dot{E}_{i}^{R}+\dot{E}_{i}^{C}+{\cal R}_{iu}\epsilon_{i}+
 {\cal R}_{i \ell}\epsilon_{\ell}
 = ({\cal R}_{i\ell}+{\cal R}_{iu})\epsilon_{i} \;\;\; .  
\end{equation}
Similarly, the probabilities that it makes an internal transition to 
{\em particular} upper or lower states are
\begin{equation}
 p_{iu}={\cal R}_{iu}\epsilon_{i}/D_{i} \;\;\;\;\; and \;\;\;\;\;
             p_{i \ell}= {\cal R}_{i \ell}\epsilon_{\ell}/D_{i} \;\;\; .
\end{equation}

	Unlike transition probabilities for real atoms, these analogues for
macro-atoms depend on ambient conditions. Consequently, in the course of
a NLTE calculation, they are iterated on just as are Eddington factors in
various other radiative tranfer schemes
(Auer \& Mihalas 1970; Hummer \& Rybicki 1971). Moreover, as with Eddington 
factors, the Monte Carlo transition probabilities are dimensionless ratios
that are likely to converge faster than do their dimensional numerators
and denominators.

\subsection{Excitation equilibrium}

When Eq.(5) is summed over all energy levels, the energy flows between
different levels cancel, giving
\begin{equation}
 \sum_{i} (\dot{A}_{i}^{R} + \dot{A}_{i}^{C}) = 
 \sum_{i} (\dot{E}_{i}^{R} + \dot{E}_{i}^{C})  \;\;\; 
\end{equation}
Thus, in statistical equilibrium, the energy stored in the form of excitation
and ionization is stationary. For the macro-atoms, this is obeyed 
rigorously by each 
activation - de-activation event since the emitted packet's energy equals
that of the absorbed packet - see Figure 1.

\section{Alternative formulations}

Monte Carlo transition probabilities have been defined in Sect. 3, but
their non-negativity was not established. Of concern in this regard is 
stimulated
emission when level populations are inverted. However, in anticipation of this
issue,
radiative rates were introduced without
specifying whether stimulated emission contributes positively to
$R_{ij}$ or negatively to $R_{ji}$. We now exploit this flexibility in order
to avoid negative probabilities.

\subsection{General case}
 
In the general case, inverted level populations may occur  - i.e.,
$g_{j}n_{i} > g_{i}n_{j}$ for some $i > j$ .

\subsubsection{Definitions of rates}

In order to prevent the probabilities becoming negative 
when levels invert, stimulated emissions must be added
to spontaneous emissions and {\em not} treated as negative absorptions.
Accordingly,
for bound-bound (b-b) transitions, the radiative rates per unit volume are
defined to be 
\begin{equation}
 R_{ij} = (A_{ij}+B_{ij} \bar{J}^{e}_{ij}) n_{i}  \;\;\;\; and
\;\;\;\;  R_{ji} = B_{ji} \bar{J}^{a}_{ji} n_{j}  \;\;\; ,
\end{equation}
where $\bar{J}^{e}_{ij}$ and $\bar{J}^{a}_{ji}$ are the mean
intensities averaged over the line's emission and absorption profiles
- see Mihalas (1978, p78). Similarly, for free-bound (f-b) and 
bound-free (b-f) transitions, we define
\begin{equation}
 R_{\kappa i}=(\alpha_{i}^{sp}+\alpha_{i}^{st}) n_{\kappa}n_{e}  \;\;\; and
\;\;\; R_{i \kappa}= \gamma_{i} n_{i}  \;\;\; .
\end{equation}
Here $\alpha_{i}^{sp}$ and $\alpha_{i}^{st}$ are the rate coefficients for 
spontaneous and stimulated recombinations to level $i$,
and $\gamma_{i}$
is the uncorrected rate coefficient for photoionizations from level $i$. Each
of these
three quantities can be expressed as an integral over frequency involving
the b-f absorption coefficient for an atom excited to level $i$ -
see Mihalas (1978, pp130-131).

        For collisions, a population inversion gives a negative rate if
de-excitations are treated as negative excitations.
This is avoided by defining
\begin{equation}
 C_{ij}=q_{ij}n_{i}n_{e} \;\;\;\;and\;\;\;\;  C_{ji}=q_{ji}n_{j}n_{e} \;\;\; .
\end{equation}

        With these expressions for the radiative and collisional rates,
the probabilities defined by Eqs.(7) and (9) are non-negative 
provided only that the $\epsilon_{i}$'s are non-negative. This latter
condition
is satisfied with the standard convention that the ground state of the 
neutral atom has zero excitation energy.

\subsubsection{Absorption of packets}

        Because $R_{\ell i}$ and therefore $\dot{A}_{i}^{R}$ are here
defined without correcting for stimulated emission, the macroscopic
line- and continuum-absorption coefficients that determine the flight paths
of $r$-packets must also be defined without this
correction. This ensures a positive absorption 
coefficient even for a transition with a population inversion.

\subsubsection{Emission of packets}

	If the Monte Carlo transition probabilities result in a macro-atom
de-activating radiatively from state $i$, the next step is to determine the
frequency of the photons comprising the emitted $r$-packet. First we
suppose that $i$ corresponds to a bound level.  
   
        Because $R_{i \ell}$ and therefore $\dot{E}_{i}^{R}$ here
include stimulated emission, the process that radiatively de-activates 
the macro-atom may be either a spontaneous or a stimulated
emission. The ratio of the probabilities of these alternatives is 
$q = \dot{E}^{sp}_{i}/\dot{E}^{st}_{i}$, where 
\begin{equation}
 \dot{E}^{sp}_{i} = A_{i \ell}n_{i} \epsilon_{i \ell} = \dot{E}^{sp}_{i \ell}
   \;\; and \;\;
 \dot{E}^{st}_{i} = B_{i \ell} \bar{J}^{e}_{i \ell}n_{i} \epsilon_{i \ell}
=  \dot{E}^{st}_{i \ell}
\end{equation}
are the contributions to $\dot{E}^{R}_{i}$ from spontaneous and stimulated
emissions. Knowing $q$, we can choose between the
two alternatives with a standard Monte Carlo procedure. Thus, if $x$ is a
random number from the interval $(0,1)$, we select spontaneous emission if 
$x < q/(1+q)$ and stimulated otherwise.  

	Having thus decided the emission process,
we must next choose a downward transition.
For spontaneous line emission, the transition    
$i \rightarrow j$ is selected with probability 
$\dot{E}^{sp}_{ij}/\dot{E}^{sp}_{i}$. For 
stimulated emission, on the other hand, the selection probability is 
$\dot{E}^{st}_{ij}/\dot{E}^{st}_{i}$.

	With the transition thus determined, the frequency $\nu$ of the
$r$-packet is selected by randomly sampling the line's
emission profile $\phi^{e}_{\nu}$. Thus, if $x$ again denotes a random number
from $(0,1)$, then $\nu$ is determined by the equation
\begin{equation}
 \int_{0}^{\nu} \phi^{e}_{\nu}\: d\nu = x  \;\; .
\end{equation}
This equation can of course always be solved numerically for $\nu$.
However, elegant and  
efficient procedures for sampling standard profiles are available
(Lee 1974a,b).

	Now we consider a macro-atom that de-activates from a
continuum state $\kappa$. In this case, the probabilities of spontaneous
and stimulated emission are in the ratio 
$\dot{E}^{sp}_{\kappa}:\dot{E}^{st}_{\kappa}$, where 
\begin{equation}
 \dot{E}^{sp}_{\kappa} = \alpha^{sp}_{i \ell}n_{\kappa} \epsilon_{\kappa \ell}  \;\;\;\;    and \;\;\;\;
 \dot{E}^{st}_{\kappa} = \alpha^{st}_{i \ell}n_{\kappa} \epsilon_{\kappa \ell}
\end{equation}
are the contributions to $\dot{E}^{R}_{\kappa}$ from spontaneous and
stimulated
emissions. Thus $\nu$ is 
selected by first deciding between spontaneous
and stimulated emission and then randomly sampling the energy distribution of
the chosen process's recombination continua.

\subsubsection{Direction of propagation}

	If the above selection procedure rules that an $r$-packet 
is emitted spontaneously, then a new direction of propagation is
chosen in accordance with this process's isotropic emission. On the other 
hand, for stimulated emission, the new direction of propagation is that of
the stimulating photon. Thus, the new direction will be in solid angle
$d \omega$ at $\theta, \phi$ with probability
$d \omega/4 \pi \times  I_{\nu} (\theta, \phi)/J_{\nu}$, where $\nu$ is the
frequency of the
emitted $r$-packet. Accordingly, a Monte Carlo code that
treats stimulated emission separately must store a complete
description of the radiation field - i.e., $I_{\nu}(\theta, \phi)$.

\subsection{Standard case}

	For problems where population inversions are not anticipated, we can
usefully make the 
traditional assumption that lines have identical emission and absorption
profiles and treat stimulated emissions as negative absorptions - see
Mihalas (1978, p78).

\subsubsection{Definitions of rates}

The radiative rates for b-b transitions are then 
\begin{equation}
 R_{ij} = A_{ij} n_{i}  \;\;\;\; and
\;\;\;\;  R_{ji} = (B_{ji}n_{j}-B_{ij}n_{i})\bar{J}_{ji} \;\;\; .
\end{equation}
Similarly, for f-b and b-f transitions, we define   
\begin{equation}
 R_{\kappa i}=\alpha_{i}^{sp} n_{\kappa}n_{e}  \;\;\; and
\;\;\; R_{i \kappa}= \gamma_{i}^{corr} n_{i}  \;\;\; ,
\end{equation}
where the photionization coefficient is now corrected for stimulated
recombinations.

	For collisions, the absence of population inversions allows us to
treat de-excitations as negative excitations without the risk
that Eqs.(7) and (9) will give negative probabilities. Accordingly,
we now define
\begin{equation}
 C_{ij}=0 \;\;\;\; and
\;\;\;\;  C_{ji} = (q_{ji}n_{j}-q_{ij}n_{i})n_{e}  \;\;\;  .      
\end{equation}
This then implies that $\dot{E}^{C}_{i}$ and therefore also $p^{C}_{i}=0$ 
for all $i$. Energy transfer from the radiation field to the thermal pool
then occurs {\em explicitly} only via f-f absorptions.

\subsubsection{Absorption of packets}

	Because $R_{\ell u}$ and therefore $\dot{A}_{i}^{R}$ are here
defined with the correction for stimulated emission included, the
macroscopic
line- and continuum-absorption coefficients  
must also include this
correction. In the posited absence of population inversions, 
these absorption coefficients are positive.

\subsubsection{Emission of packets}

	Because $R_{i \ell}$ and therefore $\dot{E}_{i}^{R}$ now
exclude stimulated emission, the process that radiatively de-activates 
a macro-atom is always a spontaneous emission. If $i$ is a bound state,
the frequency $\nu$ of the emitted $r$-packet is then decided
as follows: the transition $i \rightarrow j$
is selected with probability
$A_{ij}n_{i} \epsilon_{ij}/ \dot{E}_{i}^{R}$, and then $\nu$ is selected
by randomly sampling this transition's emission profile, as in Sect. 4.1.3.
 
	For de-activation from a continuum state, $\nu$ is
selected by randomly sampling the energy distribution of the spontaneous
recombination continua. 

\subsubsection{Direction of propagation}

Because the de-activating process is in this case spontaneous emission,
the new direction of propagation is selected according to isotropic emission.
Thus, we now do not need 
to store $I_{\nu}(\theta, \phi)$. In fact, from the Monte Carlo radiation
field generated at one iteration, we only require the mean 
intensities
$J_{\nu}$. These allow us to compute transition probabilities from
Eqs.(7) and (9) for use during the next iteration.

\subsection{Large velocity gradients}  

The procedures described in Sects. 4.1 and 4.2 apply to both static and
moving media. But for some important problems involving moving media, a
substantial speeding up of the calculation with negligible loss
of accuracy is possible by applying Sobolev's theory of line formation. In
doing so, we take advantage of a small dimensionless quantity - the ratio of
a line's Doppler width to the typical flow velocity, which implies an
essentially constant velocity gradient over the zone in which a given pencil
of radiation interacts with a particular line. The Monte Carlo codes for
hot star winds and SNe cited in Sect. 1 treat line formation in the Sobolev
approximation.

	The simplest case of this kind is that of homologous spherical
expansion, as is commonly assumed for SNe. This case will be treated
here since it will be used in the test calculations of Sect. 5.
But generalization to a spherically-symmetric stellar wind is readily
carried out by referring to Castor \& Klein (1978). We also assume 
no population inversions and so treat stimulated emissions as negative
absorptions, as in Sect. 4.2.

\subsubsection{Definitions of rates}

The radiative rates for b-b
transitions are then
\begin{equation}
 R_{ij} = A_{ij}\beta_{ji} n_{i}  \;\;\; and
\;\;\;  R_{ji} = (B_{ji}n_{j}-B_{ij}n_{i})\beta_{ji} J_{ji}^{b} \;\; .
\end{equation}
Here $J_{ji}^{b}$ is the mean intensity at the far blue wing  of the
 transition
$j \rightarrow i$, and $\beta_{ji}$ is the Sobolev escape probability
for this transition, given by
\begin{equation}
 \beta_{ji}= \frac{1}{\tau_{ji}}[1-\exp(-\tau_{ji})] \;\;\; ,
\end{equation}
where $\tau_{ji}$, the transition's Sobolev optical depth, is
\begin{equation}
 \tau_{ji}= (B_{ji}n_{j}-B_{ij}n_{i})\frac{hct_{E}}{4\pi}  \;\;\; ,
\end{equation}
with $t_{E}$ being the elapsed time since the SN exploded. For f-b and b-f
transitions, the rates are as in Eq.(18). For
collisions, the rates are as in Eq.(19).

\subsubsection{Absorption of packets}

The absorption of $r$-packets by lines is determined by the Sobolev optical
depths given by Eq.(22). Absorption of an $r$-packet to the continuum
is determined
by the conventional macroscopic absorption coefficient corrected for
stimulated emission.

\subsubsection{Emission of packets}

The frequency of an emitted $r$-packet is decided
as follows: for de-activation from a bound state $i$, the transition
$i \rightarrow j$ is selected with probability
$A_{ij}\beta_{ji}n_{i} \epsilon_{ij}/ \dot{E}_{i}^{R}$, where
$\dot{E}_{i}^{R}$ is evaluated with Eq.(1) using the decay rates from
Eq.(20), and the emitted packet is assigned frequency $\nu_{ij}^{-}$ - i.e.,
it is in the far red wing of a line whose emission profile is approximated by
a delta function. The packet's next possible b-b transition is therefore 
with the next line to the redward of $\nu_{ij}$ (Abbott \& Lucy 1985).  

	For de-activation from a continuum state, the new
frequency is, as in Sect. 4.2.3,
selected by randomly sampling the energy distribution of the spontaneous
recombination continua. 

\subsubsection{Direction of propagation}

If an $r$-packet is emitted from a continuum state, the new direction of
propagation is selected according to isotropic emission since the emission
in this case is spontaneous. For de-activation from a bound state, the
emission is also isotropic since, for homologous expansion, there
is no kinematically-preferred direction. This is not true for a stellar wind.

\section{Convergence tests}

The Monte Carlo transition probabilities derived in Sect. 3 are designed
to reproduce asymptotically the emissivity of an
atomic species whose level populations are in statistical equilibrium.
To test this, we now consider
one-point problems with specified and fixed ambient conditions.
Such tests sensibly 
precede application to a general NLTE problem, for then the
local ambient conditions are everywhere being adjusted iteratively as the
global solution is sought.

\subsection{Fe II}

In the initial tests, the Monte Carlo transition probabilities are applied
to the
model Fe II ion with $N = 394$ levels used previously (Lucy 1999b) to
investigate
the accuracy of approximate treatments of line formation in SNe envelopes.
The energy levels of the Fe II ion and the f-values for permitted
transitions were extracted from the Kurucz--Bell (1995) compilation by
M.Lennon (Munich). Einstein A-values for forbidden transitions are from
Quinet et al.(1996) and Nussbaumer \& Storey (1988). Collision strengths,
needed for Sect.5.1.5,
are from Zhang \& Pradhan (1995) and van Regemorter (1962).

\subsubsection{Radiative excitation}

	In the first Fe II test, we neglect collisional excitations
and,
as previously (Lucy 1999b), take the ambient radiation field determining the
quantities
$J_{ji}^{b}$ in Eq.(20) to be $W B_{\nu}(T_{b})$ with
$T_{b} = 12500K$ and dilution factor $W = 0.5$, corresponding to $r = R$.
The density
parameter is $n(Fe II) = 6.6 \times 10^{7} cm^{-3}$,
and the time since explosion is $t_{E} = 13\: days$. With parameters
specified, this one-point statistical equilibrium problem - Eq.(4)   
for $N-1$ levels plus a normalization constraint - is non-linear
in the unknowns $n_{i}$ because the rate coefficients in Eq.(20)
depend on the $n_{i}$ through the Sobolev escape probabilities.
Fortunately,
repeated back substitutions give a highly accurate solution $n_{i}^{(x)}$
in $\sim$ 10 iterations. 

\subsubsection{Monte Carlo experiment}

	With $n_{i}^{(x)}$ determined, the Fe II level
emissivites $\dot{E}_{i}^{R}$ and absorption rates $\dot{A}_{i}^{R}$ can be
computed from Eq.(1). We now test the Monte Carlo transition
probabilities by seeing how accurately they reproduce these values 
$\dot{E}_{i}^{R}$.
Note that it is sufficient to test {\em level} emissivities since if these
are exact
so also are the line emissivities computed as described in Sect. 4.3.3.

	In the following Monte Carlo experiment, $\cal N$ packets of
radiant energy are absorbed and subsequently emitted by a macro-atom
representing a macroscopic
volume element of Fe II ions in the ambient conditions specified above.
The energies of these packets are taken to be equal and given by
$\epsilon_{0}=\dot{A}^{R}/{\cal N}$, where
$\dot{A}^{R} = \sum_{i} \dot{A}_{i}^{R}$.
The calculation proceeds step-by-step as follows:

	1) ${\cal N}_{i} = {\cal N} \dot{A}_{i}^{R}/\dot{A}^{R}$ of the
packets activate the macro-atom to internal state $i$.

	2) The transition probabilities $p_{i}^{R}$, $p_{iu}$ and
$p_{i \ell}$ for a macro-atom in state $i$ are computed from Eqs.(7)
and (9).

	3) The transition probabilities sum to one, so each corresponds
to a segment $(x_{k},x_{k+1})$ of the interval (0,1). A particular
transition is therefore selected
by computing a random number $x$ in (0,1) and finding in which segment it
falls.

	4) If the selected transition is the de-activation of the
macro-atom, we update $\dot{E}_{i}^{MC}$ to $\dot{E}_{i}^{MC}+\epsilon_{0}$
and then return to step 3) to process the next activation of state $i$, 
or to step 2) to process the first of the packets
that activate the macro-atom to state $i+1$.    

	5) If the selected transition is an internal transition
to state $j$, then we return to step 2) with $j$ replacing $i$. 

	6) When all $\cal N$ packets have been processed, the
final elements
of the vector $\dot{E}_{i}^{MC}$ are the estimates of the level emissivities 
$\dot{E}_{i}^{R}$

\subsubsection{Results of experiment}

	As a single measure of the accuracy of the estimated level
emissivities, we compute the quantity
\begin{equation}
 \delta = \sum_{i} |\dot{E}_{i}^{MC}-\dot{E}_{i}^{R}|\;/
 \sum_{i} \dot{E}_{i}^{R}  \;\;\; .
\end{equation}	
This is the mean of the absolute fractional errors of the 
$\dot{E}_{i}^{MC}$ when weighted by $\dot{E}_{i}^{R}$.

	Figure 2 shows the values of $\delta$, expressed as percentage
errors, found in a series of trials with $\cal N$ increasing from $10^{4}$ to
$10^{7}$. The values of $\delta$ decrease monotonically with increasing
$\cal N$, falling to $0.36$ percent for ${\cal N} = 10^{7}$. More importantly,
the errors accurately follow an ${\cal N}^{-1/2}$ line, as expected if the
only source of error are the sampling error at step 3) of the
Monte Carlo experiments. Accordingly, to the accuracy of these experiments,
macro-atoms obeying the transition probabilities derived in Sect. 3 do
indeed reproduce the emissivity of a gas in statistical equilibrium.

\begin{figure}
\vspace{8.2cm}
\includegraphics{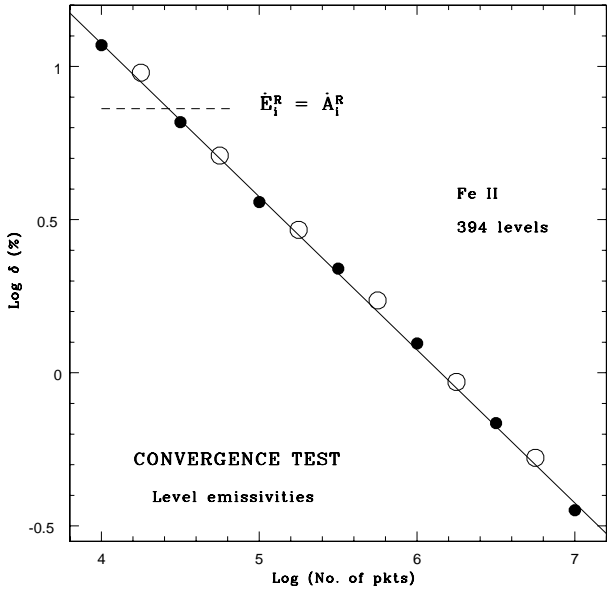}
\caption{Convergence test. The mean error $\delta$ defined by Eq.(23)
is plotted against $\cal N$, the number of packets in the Monte Carlo 
experiment. The open circles refer to the case where excitation energies are
increased by 5eV. The straight line drawn by eye has slope $= - 0.5$. Also
indicated is
the mean error when the level emissivities are assumed equal to the level
absorption rates.}
\end{figure}

	Also included in Fig.2 are values of $\delta$ obtained when the
transition probabilities are computed with excitation energies 
$\epsilon_{i}$ increased by 5eV. This is to investigate the consequences
of the dependence of the energy flow terms in Eq.(5) - and therefore
also of the transition probabilities - on the zero point of the scale of
excitation
energy. These results also track an ${\cal N}^{-1/2}$ line and so indicate
that the predicted emissivities are asymptotically independent of the
zero point. But since the open circles are marginally higher, there is an 
indication that increasing the zero point gives slighty less accurate
emissivities at a given $\cal N$.

	In the Monte Carlo codes for hot star winds and SNe cited in Sect. 1,
line formation is treated approximately, with either resonant scattering
or downward branching being assumed. For both
assumptions,
$\dot{E}_{i}^{R} = \dot{A}_{i}^{R}$, corresponding to a macro-atom for which
de-activation always immediately follows activation - i.e., $p_{i}^{R} = 1$
for all $i$. In this case, as indicated on Fig.2, $\delta = 7.28$ percent.
Thus, when the points in Fig.2 drop below this value, the success must be
due to the internal, radiationless transitions governed by the probabilities 
$p_{iu}$ and $p_{i \ell}$.   

\subsubsection{Distribution of jumps}

	The above experiments show that despite the formidable complexity of
its level
structure the Fe II ion's reprocessing of radiation is accurately simulated
by the Monte Carlo transition probabilities. Nevertheless, from a
computational
standpoint, a remaining concern is how many internal 
transitions - or jumps - does this require? To answer this, 
the number of jumps before de-activation was recorded for each absorbed
packet in the ${\cal N} = 10^{7}$ trial and used to derive $N(j)$, the
number of packets requiring $j$ jumps.

  	From $N(j)$, we find that the expected
number of
jumps is $<\!j\!> \:= 2.19$ and that the probability of immediate
de-activation -
i.e., zero jumps - is $P_{0} = 0.425$. Evidently, fears of numerous, time-
consuming internal transitions are ill-founded.

\begin{figure}
\vspace{8.2cm}
\includegraphics{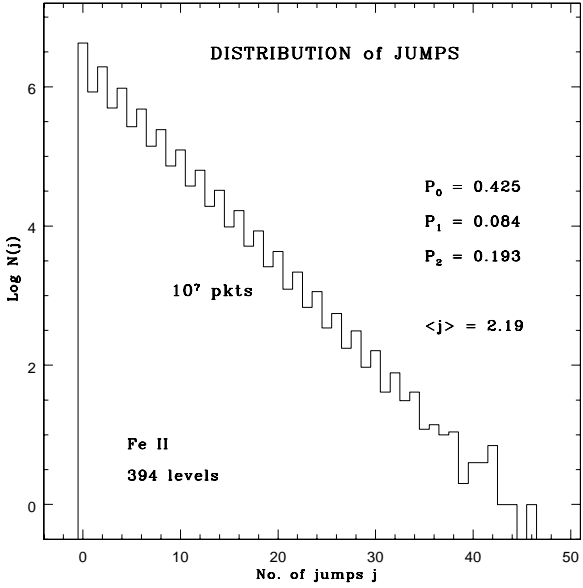}
\caption{Histogram of $N(j)$, the number of times in an experiment with 
${\cal N} = 10^{7}$ that the macro-atom underwent
$j$ internal transitions - or jumps - before de-activating with the emission
of an energy
packet. The mean number of jumps $<j>$ and the probabilities of de-activation
after $j = 0-2$ jumps are indicated.
}
\end{figure}

	Figure 3 is a logarithmic plot of $N(j)$. This reveals a power-law
decline with increasing $j$ but with alternating deviations
indicating that an even number of jumps before de-activation
is favoured. A simple model suggests the origin of this curious behaviour.
Consider a 3-level atom with $\epsilon_{3} > \epsilon_{2} > \epsilon_{1} = 0$
and suppose that level 2 is metastable with $A_{21} = 0$. Because
$B_{12} = 0$, the macro-atom can only be activated to state 3; and because
$A_{21} = B_{21} = 0$, the macro-atom cannot de-activate from state 2.
Moreover,
since $\epsilon_{1} = 0$, Eq.(9) gives $p_{31} = p_{21} = 0$, and so
state 1 of the macro-atom cannot be reached. Accordingly, following
activation at state 3,
the macro-atom de-activates with probability $p$ or jumps to state 2 with
probability $1-p$, from whence it returns to state 3 with probability
$p_{23}=1$. It is now simple to prove that the probabilty of $j$ jumps 
before de-activation is
$P_{j}=p(1-p)^{j/2}$ if $j$ is even, and $P_{j}=0$ if $j$ is odd. The Fe II
ion's numerous low-lying metastable levels are presumably playing the role
of level 2 and thereby favouring an even number of jumps.  

	Histograms $N(j)$ have also been computed for two other cases. First,
the above trial was repeated with the $\epsilon_{i}$'s increased by 5eV as in
Sect. 5.1.3. This change increases $<j>$ - to 4.54 - as expected since the
probabilities $p_{iu}$ and $p_{i \ell}$ are thereby increased and 
$p_{i}^{R}$ correspondingly decreased. Evidently, the
standard choice of energy-level zero point leads to the most
computationally-efficient set of transition probabilities.

	In the second case, $W$ is decreased from 0.5 to
0.067, corresponding to $r = 2R$. This change decreases $<j>$ - from 2.19 to
1.29 -  as expected given the weakening of the radiative excitation rates.

\subsubsection{Collisional excitation}

	In the above experiment, the emission derives entirely
from radiative excitation since collisions were neglected. Now we
consider the opposite extreme by setting the ambient radiation field to zero
but including collisions. 

	The only parameters of this test are the electron temperature and 
density, and these are assigned the values $T_{e} = 2 \times 10^{4} K$ and
$N_{e} = 10^{8} cm^{-3}$. The resulting statistical equilibrium 
problem is linear and so solved without iteration. For this solution,
accurate values of the level emissivities $\dot{E}_{i}^{R}$ are again computed
from Eq.(1). 

	The next step is to derive estimates of the level
emissivities by repeating the Monte Carlo experiment of Sect.5.1.2. The
only changes needed are the following: first, since the solution has
population
inversions the general formulation of Sect. 4.1 must be adopted to avoid
negative probabilities.

	 Secondly, since a macro-atom is now
always activated by a $k$-packet, their energies are
taken to be $\epsilon_{0}=\dot{A}^{C}/{\cal N}$, where
$\dot{A}^{C} = \sum_{i} \dot{A}_{i}^{C}$. Correspondingly, at
step 1) of the experiment,
${\cal N}_{i} = {\cal N} \dot{A}_{i}^{C}/\dot{A}^{C}$.	
	
	Thirdly, since a macro-atom can now de-activate by emitting either an
$r$- or a $k$-packet, only the former results in an updating of
$\dot{E}_{i}^{MC}$. The emission of a $k$-packet represents the return of
energy $\epsilon_{0}$ to the therrmal pool.

	Apart from these changes, the convergence experiment proceeds as
in Sects. 5.1.2 and 5.1.3. The result is a plot similar to Fig.2, but with
$\delta = 0.19$ percent for ${\cal N} = 10^{7}$. Evidently, the Monte Carlo 
transition probabilities are equally applicable to problems where
collisional excitation is a source of emission.

\subsection{Hydrogen}

	Although the Fe II experiments demonstrate the validity
of the Monte Carlo transition probabilities, a test including b-f and f-b
transitions is of interest. Accordingly, a convergence experiment at
one point in
a SN's envelope has also been carried out for a 15-level model of the
H atom, with
level 15 being the continuum $\kappa$. The 14 bound levels correspond to 
principal quantum numbers $n = 1-14$, with each level having consolidated
statistical weight $g = 2n^{2}$.

	As for Fe II, the ambient radiation field incident on the
blue wings of the b-b transitions is $WB_{\nu}(T_{b})$, but now with 
$T_{b} = 6000K$ and $W = 0.067$. However, beyond the Lyman limit, we assume
zero
intensity, so that photoionizations occur only from excited states.
Correspondingly, recombinations to $n=1$ are excluded on the assumption of
immediate photoionization. Collisional excitations and ionizations are
neglected. The density parameter is
$N(H) = 1.88 \times 10^{9} cm^{-3}$, the electron temperature
$T_{e} = 4800K$, and the time since explosion
$t_{E} = 10 \: days$. 
With parameters specified, this non-linear statistical equilibrium problem 
can also be solved with repeated back substitutions, giving a highly accurate
solution $n_{i}^{(x)}$ in $\sim 30$ iterations.

	With  $n_{i}^{(x)}$ determined, Monte Carlo experiments as described 
in Sect. 5.1.2 were carried out to test if level emissivities are also 
recovered
in this case. In Fig.4, two such trials, with ${\cal N} = 10^{4}$ and
$10^{5}$,
are compared with the exact solution. The results show that
excellent agreement is achieved for ${\cal N} = 10^{5}$. Note in particular
the success
with $\dot {E}_{\kappa}^{R}$, which is the rate of ionization energy loss
due to recombinations, and with $\dot {E}_{2}^{R}$, whose very low
value is due to the strong trapping of $L \alpha$ photons.

\begin{figure}
\vspace{8.2cm}
\includegraphics{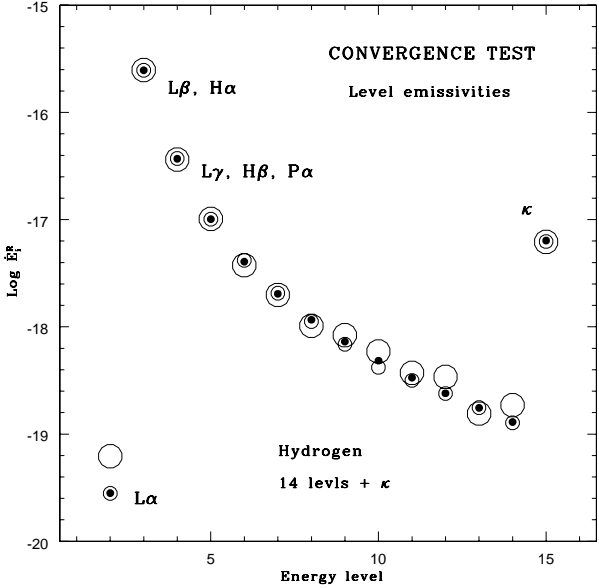}
\caption{Level emissivities (cgs) for Hydrogen. Results for trials with
${\cal N} = 10^{4}$ (large open circles) and ${\cal N} = 10^{5}$ (small
open circles) are compared with exact values (filled circles). The lines
contributing to the level emissivities are indicated for $n = 2-4$.}
\end{figure}

\subsection{Alternative test of convergence}

	Thus far, a Monte Carlo procedure has been used to
validate the transition probabilities developed in Sect.3. This
has the advantage of following closely and therefore illustrating their use
in realistic NLTE calculations. But for feasible values of $\cal N$, sampling
errors limit the accuracy of such tests.

	In order to test to higher precision, approximate level emissivities  
$\dot {E}_{i}^{(m)}$ can be computed recursively according to the following
scheme:
\begin{equation}
\dot {E}_{i}^{(m)} = p_{i}^{R}  \sum_{r=1}^{m} G_{i}^{(r)}   \;\;\;  ,
\end{equation}	
where $p_{i}^{R}$ is the radiative de-activation probability from Eq.(7) and
$G_{i}^{(r)}$ is the increment at cycle $r$ to the summation approximating the
rate at which level $i$ gains energy - i.e. the right-hand side of
Eq.(5). This increment is derived from the previous increment by applying
the transition probabilities from Eq.(9). Thus 
\begin{equation}
 G_{i}^{(r)} =  \sum_{j} p_{ji} G_{j}^{(r-1)}   \;\;\;  ,
\end{equation}	
and the recursion cycles are initiated by setting
\begin{equation}
 G_{i}^{(1)} = \dot {A}_{i}^{R} + \dot {A}_{i}^{C}   \;\;\;  .
\end{equation}	

	This procedure is now applied to the Fe II test problem of
Sect. 5.1.1. As with that experiment, the accuracy of the vectors
$\dot {E}_{i}^{(m)}$ are measured by computing  $\delta$ defined by Eq.(23).
For $m=17$, $\delta$ drops below the value 0.36 percent found in Sect.5.1.3
with ${\cal N} = 10^{7}$ - see Fig.2. As the recursion procedure continues
further,
$\delta$
decreases monotonically until at $m \simeq 60$ it drops to a value of
$\simeq 10^{-8}$, at 
which point machine precision or accumulated roundoff errors halt further
progress. This test clearly confirms and strengthens the earlier tests of the
Monte Carlo transition probabilities.

\section{Sensitivity tests}

The experiments of Sect.5 demonstrate that, when computed with the exact level
populations $n_{i}^{(x)}$, the Monte Carlo transition probabilities
applied to indivisible $e$-packets reproduce the exact level emissivities
as ${\cal N} \rightarrow \infty$. But this
success, though necessary, does not of itself imply that the technique will
be successful when applied to NLTE problems. For example, if the Monte Carlo
emissivities were to
undergo large changes in response to small changes in $n_{i}$,
then we would reasonably suspect that the iterations inevitably
required for a NLTE
problem would converge very slowly - or might even diverge. On the
other hand, if the
emissivities are insensitive to changes in $n_{i}$, then the
prospects for successful applications are excellent. 

\subsection{Fe II emissivities}

This crucial question of sensitivity can be investigated by 
repeating the calculations of Fe II emissivities reported in 
Sect.5.1, but with $n_{i}$ perturbed away from $n_{i}^{(x)}$. A convenient
way of doing this is to replace $n_{i}^{(x)}$ by the
Boltzmann distribution at excitation temperature $T_{ex}$. 
Then, for given $T_{ex}$, the corresponding level emissivities
$\dot{E}_{i}^{MC}$
are obtained from a Monte Carlo trial with ${\cal N} = 5 \times 10^{6}$
packets, and so are negligibly affected by sampling errors (cf. Fig.2) .

	Now, for the given $T_{ex}$, we can also
compute $\dot{E}_{i}^{R}$, the level emissivities predicted by the 
fundamental formulae - Eqs.(1) and (20) in this case. This 
represents the standard approach to NLTE transfer problems whereby the
radiation field is
computed from the Radiative Transfer Eq. (RTE) with emissivity
coefficients evaluated using the current estimates of $n_{i}$.
Thus by comparing these two emissivity estimates $\dot{E}_{i}^{MC}$
and $\dot{E}_{i}^{R}$, we can see whether this Monte
Carlo technique is potentally capable of yielding a superior estimate of the
radiation field.

	In Fig.5, the quantities  $\dot{E}_{i}^{MC}$ and
$\dot{E}_{i}^{R}$ obtained for $T_{ex} = 12500K$ are plotted against
$\dot{E}_{i}^{(x)}$, the exact statistical equilibrium level
 emissivities
- i.e., the values corresponding to $n_{i}^{(x)}$.

\begin{figure}
\vspace{8.2cm}
\includegraphics{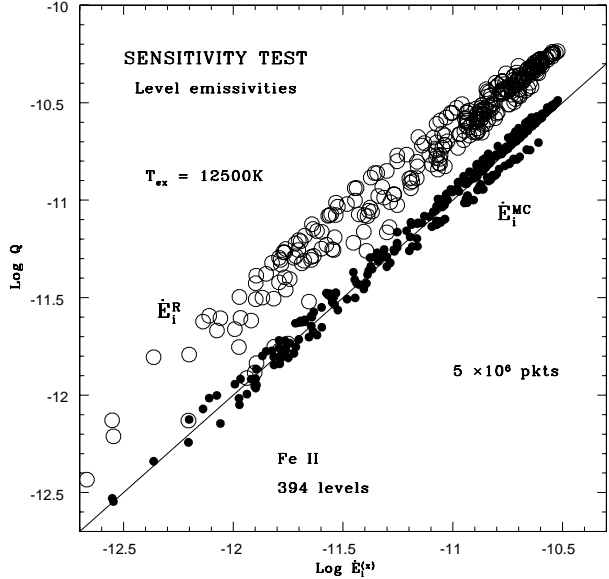}
\caption{Sensitivity test. For a Boltzmann distribution over excited states 
at $T_{ex}=12500K$, the level emissivities (cgs) obtained with the
Monte Carlo transition probabilities (filled circles) and with the basic
formula (open circles) are plotted against the exact emissivities obtained
with $n_{i}^{(x)}$. The Monte Carlo emissivities derive from a trial with
${\cal N} = 5 \times 10^{6}$ packets.}
\end{figure}

	Remarkably, Figure 5 shows that the Monte Carlo 
emissivities are far less sensitive to the
departure of $n_{i}$
from $n_{i}^{(x)}$ than are the emissivities computed directly from the
fundamental formula. For the most part, the $\dot{E}_{i}^{MC}$ are in error by
$< 0.1$ dex, with little evidence of bias, while the $\dot{E}_{i}^{R}$
are systematically offset by $\sim +0.3$ dex.

	To investigate whether this insensitivity is characteristic
of the Monte Carlo procedure, the above test is now repeated with $T_{ex}$
ranging
from $7500K$ to $20000K$ and the resulting mean errors defined by
Eq.(23) plotted in Fig.6. We see that  
$\dot{E}_{i}^{R}$ gives reasonably accurate emissivities 
only in the immediate neighbourhood of the minimum at
$T_{ex} \simeq 11250K$.
On the other hand, the values $\dot{E}_{i}^{MC}$ are accurate to
$\la 0.1$ dex across the entire range.

\begin{figure}
\vspace{7.8cm}
\includegraphics{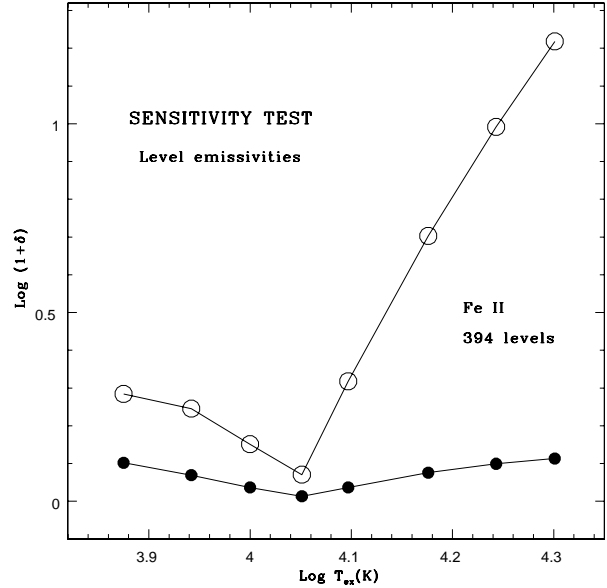}
\caption{Sensitivity test. Logarithmic errors of the emissivity vectors
$\dot{E}_{i}^{MC}$ and $\dot{E}_{i}^{R}$ evaluated for Boltzmann
distributions over excited states plotted against $T_{ex}$. The Monte Carlo
emissivities derive from trials with
${\cal N} = 5 \times 10^{6}$ packets.}
\end{figure}

	The causes of these astonishing differences in sensitivity are of
considerable interest.
For $\dot{E}_{i}^{R}$, the strong sensitivity to $T_{ex}$ is readily
understood. Because the sum
$R_{i \ell} \epsilon_{i \ell} \propto n_{i}$, an error in the 
population of the emitting level translates directly into an error in
$\dot{E}_{i}^{R}$.

	Now consider $\dot{E}_{i}^{MC}$. This quantity is
determined by the rate at which active macro-atoms reach state $i$,
and this happens by direct absorptions of packets into this state or by 
transitions from other states. Either way, the accuracy of the source vectors
$\dot{A}_{i}^{R}$ and $\dot{A}_{i}^{C}$ is clearly fundamental to the
accuracy of the vector $\dot{E}_{i}^{MC}$.
But the dominant contributors to the
elements of these source vectors - see Eqs.(1) and (2) - are
transitions from
the ground state and from low-lying metastable levels, and the estimated 
populations of these levels are unlikely to be seriously in error. In 
particular, with an assumed Boltzmann 
distribution over excited states, the $n_{i}$ of these low levels is
insensitive to $T_{ex}$ and do not differ much from $n_{i}^{(x)}$. In
contrast, the populations of high levels are quite likely to be badly
estimated and are acutely sensitive to $T_{ex}$.

\subsection{Comments}

	Another way of appreciating the differences in these approaches to
calculating emissivities is as follows. The Monte Carlo procedure applies only
to a state of statistical equilibrium and, as such, constrains every level's 
emissivity to be consistent with the rates of processes populating that
level. In
contrast, the fundamental emissivity formula applies also to states out of
statistical equilibrium and so takes no account of whether the levels'
populations 
can be maintained. Accordingly, with this Monte Carlo technique,
the principle of statistical equilibrium is {\em incorporated}
(approximately) 
as the radiation field is being calculated. On the other hand, 
when emissivities are computed from the fundamental formula, any
consideration of
statistical equilibrium is effectively being deferred until the updated 
radiation field has been determined.      

	The likely beneficial impact of this insensitivity on the iterations
needed to derive NLTE solutions is worth stressing. With the conventional
RTE approach, an erroneously overpopulated upper level $i$ pollutes the
radiation field with spurious line photons at frequencies $\nu_{ij}\; (j<i)$,
and these are sources of excitation for level $i$ when level populations are
next solved for. Similarly, an erroneously overpopulated upper ion pollutes
the radiation field with recombination photons that are subsequent sources
of photoionization for the lower ion.
To some degree, therefore, such errors are 
{\em self-perpetuating} and so are not rapidly eliminated. This persistency
contributes to the slow convergence typical of NLTE codes. In contrast, with
the Monte Carlo approach, this pollution does not happen and so - for  
sufficiently large ${\cal N}$ - a high
quality radiation field is obtained 
immediately provided that the initial populations of the low-lying levels are 
estimated sensibly.

\section{Implementation}

The Monte Carlo transition probabilities allow statistical equilibrium to be
incorporated into the calculation of radiation fields for NLTE problems.
Moreover, this is achieved without imposing the constraint of radiative
equilibrium. Accordingly, in principle at least, the technique applies
equally to problems with non-radiative heating, such as stellar
chromospheres.

\subsection{Radiative equilibrium}

In the absence of non-radiative heating, a NLTE transfer problem
must be solved subject to the constraint of radiative equilibrium. The
incorporation of this {\em additional} constraint into the macro-atom
formalism is readily understood. First suppose that collisional processes are
neglected. The absorbed and the emitted $e$-packets
are then always $r$-packets and they have identical energies
- see Fig.1. Thus, the constraint of radiative
equilibrium is obeyed rigorously since it holds exactly for every 
activation - de-activation event, all of which are of the form
$r \rightarrow {\cal A}^{*}  \rightarrow r$, where ${\cal A}^{*}$ denotes
an active macro-atom. Note also that since active macro-atoms do not appear
spontaneously within the computational domain (D), every Monte Carlo
quantum's interaction history starts and ends as an
$r$-packet crossing a boundary of D.
 
	Now suppose that collisions are included. In this case, a
macro-atom activated by an $r$-packet can
de-activate itself by emitting a $k$-packet, so that 
radiative equilibrium no longer holds exactly for each individual 
activation - de-activation event. However, the emitted $k$-packet is 
re-absorbed {\em in situ} by another macro-atom and thereby (eventually)
converted into an $r$-packet. Since this has the same energy as the original
$r$-packet, radiative equilibrium holds for every sequence of in situ events
that
starts with the absorption of an $r$-packet and ends with the next emission
of an $r$-packet. A typical in situ sequence is 
$r \rightarrow {\cal A}^{*}  \rightarrow k \rightarrow   {\cal A}^{*}  \rightarrow k  \rightarrow
{\cal A}^{*}  \rightarrow r$. If such sequences are abbreviated as  
$r \rightarrow [{\cal A}^{*}]  \rightarrow r$, we see that the inclusion of
collisions has not fundamentally changed the procedure and that radiative
equilibrium is still rigorously obeyed.

\subsection{Non-radiative heating} 

In the presence of non-radiative heating, the NLTE problem is not subject to
the additional constraint of radiative equilibrium. Statistical equilibrium
is incorporated with the macro-atom formalism as before, and the challenge
now is to incorporate the creation of radiant energy within D due
to the additional heating. This is accomplished by allowing for the
spontaneous
and random appearance within D of active macro-atoms with their number,
locations and
internal states $i$ all controlled by the collision source vector
$\dot{A}_{i}^{C}$ - 
cf. Sect.5.1.5. Note that because this sampling of $\dot{A}_{i}^{C}$ takes
full account of the collisional
creation of excitation, the emission of
a $k$-packet is not now followed by its in situ re-absorption; instead,
the interaction
history of that Monte Carlo quantum then ends and
its energy is added to the thermal pool (cf. Sect.5.1.5.).
The radiation
field generated by this procedure is not divergence-free but reflects
the collisional creation of radiant energy due to an elevated temperature
profile maintained by the non-radiative heating.

\section{Conclusion}

The limited aim of this paper has been to see if Monte Carlo transfer codes 
whose quanta are indestructable energy packets can be constructed without
resorting to 
simplified treatments of line formation. To this end, the concept
of a macro-atom has been introduced and rules established governing its
activation and de-activation as well as its transitions between internal
states. These rules - the Monte Carlo transition probabilities - have been
derived by demanding that the macro-atom's emission of $r$-packets
asymptotically reproduces the local
emissivity of a gas in statistical equilibrium; and these
rules' validity has been confirmed with one-point test problems.

	Evidently, the next step is to implement these transition
probabilities in a code to solve a realistic NLTE problem for a stratified
medium and thus to investigate the practicality of this technique for
problems of current interest. In a companion paper, a Monte
Carlo NLTE code treating the formation of H lines in a Type II SN envelope
will be described and used to illustrate the convergence behaviour of
iterations to obtain both the level populations and the temperature
stratification.

\begin{acknowledgement}
I am grateful to C.Jordan for her interest in the potential application of
this technique to stellar chromospheres as well as for a constructive
referee's report.
\end{acknowledgement}

\end{document}